# Hidden domain boundary dynamics towards crystalline perfection

A. Mangu[1,2], V. A. Stoica[3,4], H. Zheng[4,5], T. Yang[3], M. Zhang[3], H. Wang[2,3], Q. L. Nguyen[6], S. Song[6], S. Das[7,8,9], P. Meisenheimer[7,8], E. Donoway[10], M. Chollet[6], Y. Sun[6], J.J. Turner[2,6], J. W. Freeland[4], H. Wen[4,5], L. W. Martin[7,8,11], L.-Q. Chen[3], V. Gopalan[3], D. Zhu[6], Y. Cao[5], A. M. Lindenberg[1,2,12]

[1] *Department of Materials Science and Engineering, Stanford University, Stanford, CA 94305, USA*
[2] *Stanford Institute for Materials and Energy Sciences, SLAC National Accelerator Laboratory, Menlo Park, CA 94025, USA*
[3] *Department of Materials Science and Engineering, Pennsylvania State University, University Park, PA 16802, USA*
[4] *Advanced Photon Source, Argonne National Laboratory, Lemont, IL 60439, USA*
[5] *Materials Science Division, Argonne National Laboratory, Lemont, IL 60439, USA*
[6] *Linac Coherent Light Source, SLAC National Accelerator Laboratory, Menlo Park, CA 94025, USA*
[7.] *Materials Sciences Division, Lawrence Berkeley National Laboratory, Berkeley, CA 94720, USA*
[8] *Department of Materials Science and Engineering, University of California, Berkeley, Berkeley, CA 94720, USA*
[9.] *Materials Research Centre, Indian Institute of Science, Bangalore, 560012, Karnataka, India*
[10] *Department of Physics, University of California, Berkeley, Berkeley, CA 94720, USA*
[11] *Departments of Materials Science and NanoEngineering, Chemistry, and Physics and Astronomy and the Rice Advanced Materials Institute, Rice University, Houston, USA 77005, USA*
[12] *Stanford PULSE Institute, SLAC National Accelerator Laboratory, Menlo Park, CA 94025 USA*

## Abstract

A central paradigm of non-equilibrium physics concerns the dynamics of heterogeneity and disorder, impacting processes ranging from the behavior of glasses to the emergent functionality of active matter. Understanding these complex mesoscopic systems requires probing the microscopic trajectories associated with irreversible processes, the role of fluctuations and entropy growth, and the timescales on which non-equilibrium responses are ultimately maintained. Approaches that illuminate these processes in model systems may enable a more general understanding of other heterogeneous non-equilibrium phenomena, and potentially define ultimate speed and energy cost limits for information processing technologies. Here, we apply ultrafast single shot x-ray photon correlation spectroscopy to resolve the non-equilibrium, heterogeneous, and irreversible mesoscale dynamics during a light-induced phase transition. This approach defines a new way of capturing the nucleation of the induced phase, the formation of transient mesoscale defects at the boundaries of the nuclei, and the eventual annihilation of these defects, even in systems with complex polarization topologies. A non-equilibrium response spanning >10 orders of magnitude in timescales is observed, with multistep behavior similar to the plateaus observed in supercooled liquids and glasses. We show how the observed time-dependent long-time correlations can be understood in terms of the stochastic dynamics of domain walls, encoded in effective waiting-time distributions with power-law tails. This work defines new possibilities for probing the non-equilibrium and correlated dynamics of disordered and heterogeneous media.

## Main Text

While there have been numerous advances in approaches to drive matter into novel metastable phases and across phase boundaries, few of these approaches provide for direct visualization of the intrinsic dynamical heterogeneity and non-equilibrium response that often underlies these processes[1–3]. Non-equilibrium states are not, as often thought, always short-lived or negligible after some brief waiting time[4]. As a prototypical example, consider the process of rapidly tuning matter across a phase boundary to a symmetry-broken phase. This corresponds to an ultrafast quench in which phases nucleate and grow in a coarsening, self-similar response described by a power law, essentially enforcing non-equilibrium response and ergodicity breaking to long times.[4] Understanding these scale-free and universal processes requires experimental tools and approaches which probe the intrinsic role of fluctuations, disorder, heterogeneity, and memory of initial conditions[5–8], and not just the average structure as probed by crystallographic techniques[9]. Similarly, most classical nucleation and growth models focus on the volume fraction of the transformed phase, neglecting the role of the interface or domain walls.[10] Phenomenological models have been developed for such phenomena arising from time-dependent Landau-Ginzburg functionals, which can describe these nonlinear, self-similar dynamical responses.[11,12] But theoretical models which capture the non-equilibrium response of the defects themselves (*e.g.*, the non-equilibrium dynamics within a domain wall and their evolution under external stimuli) remain to be fully developed. From a broad perspective, defect-defect interactions (*e.g.*, domain-wall-domain-wall interactions or other types of correlated motion) are poorly understood but underlie a vast range of physics and technological applications.[13,14]

State-of-the-art time-resolved x-ray or electron scattering largely neglects these defect states and assumes a coherent response across all unit cells, or performs an effective ensemble average, thus blurring variations in the local structure and the presence of phase boundaries.[1–3,9,15] Instead, approaches capable of probing the small volume fraction associated with the boundaries of domains (*i.e.*, domain walls and other defect states) and more broadly, the microscopic trajectories that a heterogeneous material follows through configuration space as it transforms are required. X-ray photon correlation spectroscopy (XPCS) offers a method of monitoring the dynamic heterogeneity inherent in phase changes, but is most often applied to probe dynamics on much slower time-scales than investigated here[16,17]. Under coherent x-ray illumination, the temporal evolution of nano/mesoscale features gives rise to time-varying modulations to the diffraction pattern,[18,19], enabling new approaches for correlating diffraction images. From this perspective, we find here that even simple questions like “when is a phase transformation complete?” taken on new meaning. When tracking the correlation of the diffracted spot rather than the integrated intensity, our experiments indicate that a transformation is only half complete, when the conventional probe says it is nearly complete. In this dramatic difference lies hidden information about interfaces and defects that are created, their subsequent interactions, and their eventual annihilation during the transformation as the system moves towards crystalline perfection.

$(PbTiO_3)_n/(SrTiO_3)_n$ (PTO/STO) superlattices offer a platform to experiment with previously unexplored mesoscale-phase transitions because they exhibit a wide array of topological-polar orders, ranging from skyrmions to vortices and merons, with interesting mesoscale features.[20–22] Previous work probing PTO/STO superlattices has shown the existence of a long-lived metastable vortex supercrystal (VSC) phase only accessible through excitation by ultrafast pulses.[23] Figure 1a shows reciprocal space mapping in the $Q_z$-$Q_y$ plane about the pseudo-cubic 002-diffraction peak of the superlattices and the $DyScO_3$ substrates upon which they are grown before and after irradiation by a single 100 femtosecond (fs) pulse of 400 nm light, showing evidence for a solid-solid structural phase transition.

We conducted a single-shot optical pump-x-ray probe XPCS experiment at the Linac Coherent Light Source at the SLAC National Accelerator Laboratory to monitor the non-equilibrium speckle evolution during this irreversible light-induced phase transition (Methods). A sequence of 30 Hz repetition-rate, 9.8 keV probe x-ray pulses with ~50 fs pulse widths was incident on the sample (labeled in Fig. 1b by index $n$) with a single 400 nm pump pulse (50 fs duration) arriving at controllable time delays relative to one of the central x-ray pulses in the train. Thus, we obtained coherent speckle pattern snapshots of the equilibrium state before exposure ($n < 0$), the transiently evolving intermediate structure ($n = 0$), and the final state after the transition is complete ($n > 0$), which can be correlated relative to each other. Because the transformation is irreversible, after each sequence of pulses associated with a single pump pulse, the sample was translated to a new spot and the sequence was repeated. The resulting speckle correlation was obtained by averaging the correlation functions from each sequence. The orientation of the sample was such that the first-order satellite peak associated with the VSC phase emerges as the transformation progresses (Fig. 1b).

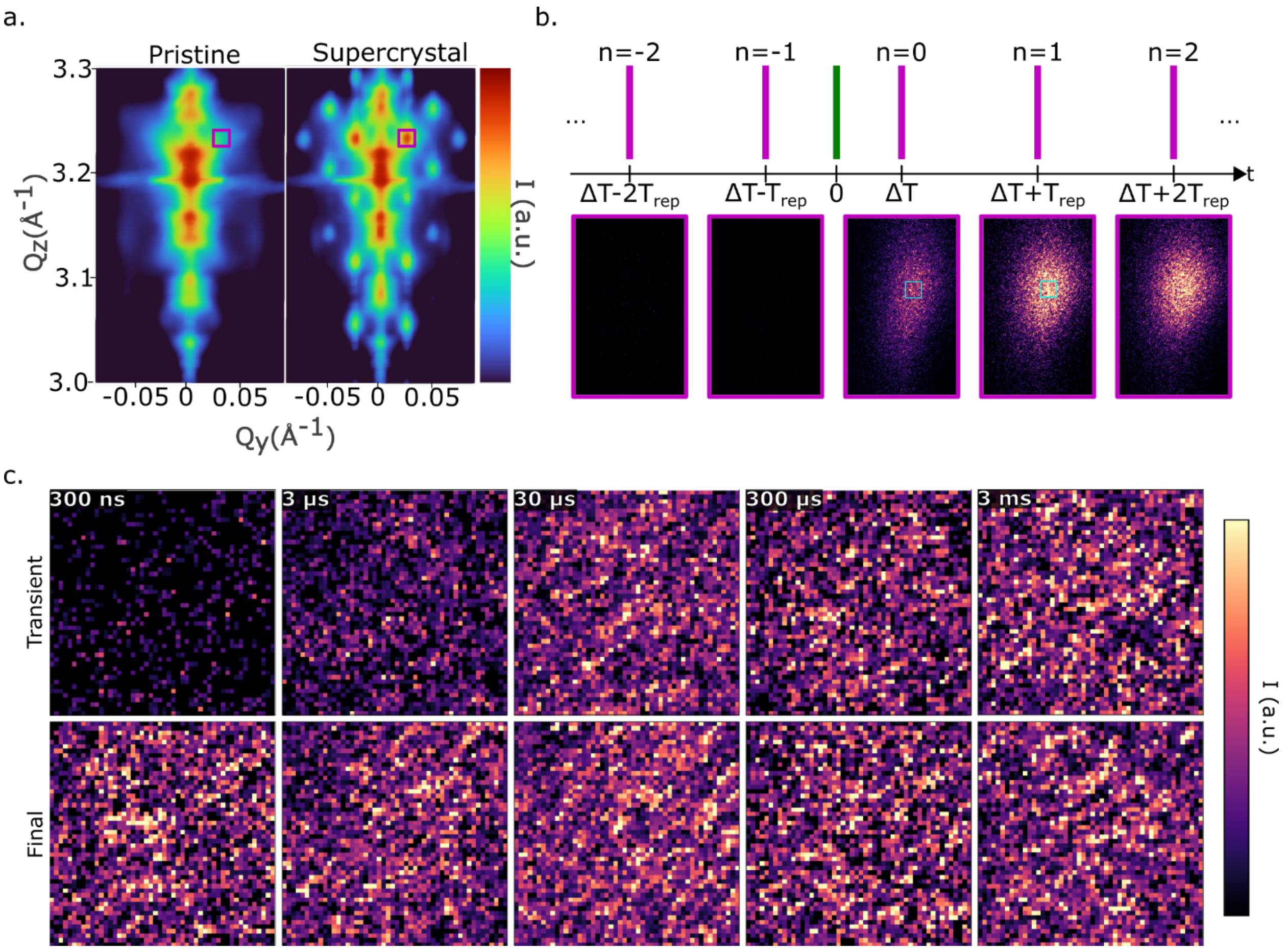

Fig. 1 a) Reciprocal space maps of pristine and transformed structures. Red box shows area of reciprocal space shown in Fig. 1b. b) Single shot XPCS setup showing the pulse-train sequence (top) and sample diffraction images (bottom) capturing pristine, transient intermediate (at ΔT = 1 μs), and final state speckle patterns corresponding to boxed area in Fig. 1a. Green boxes show the region of the satellite peak shown in Fig. 1c. c) Representative single location 50 x 50-pixel regions (centered at the peak centers extracted by procedure detailed in Methods) at the center of the transient and final supercrystal satellite peak at selected ΔT values. Whereas the integrated intensity of the entire peak saturates after tens of microseconds, the correlations in the speckle patterns continue to evolve up to millisecond timescales.

Figure 1c shows the evolution of the single shot speckle pattern as a function of time between the optical excitation pulse and the x-ray probe pulse (corresponding to $n = 0$, Fig. 1b); with the bottom row showing the final state after the transformation is complete (*e.g.*, at $t = t_\infty$). Each time corresponds to a different location on the sample and the experiment samples approximately 20 spots at each delay. Qualitatively, whereas the integrated intensity reaches its final value on timescales of order ~10 μs, correlations in the speckle pattern comparing the transient shot to the final state continue to evolve on millisecond timescales.

These speckle patterns encode information about the mesoscale structure of the induced supercrystal phase as it nucleates. We quantify the correlation of images at two different times using a pixel-by-pixel average to compute the time dependence of the correlation, using a two-time correlation function given by:[24]

$$C(t_1, t_2) = \frac{\overline{I(t_1)I(t_2)}}{\overline{I(t_1)}\,\overline{I(t_2)}} \tag{1}$$

where $I$ is the scattering intensity captured on a two-dimensional detector in a region of reciprocal space, $t_1$ and $t_2$ denote the times x-ray probe pulses arrive at the sample relative to the optical pump pulse, and the bars indicate an average taken over a region of interest (ROI) in momentum ($Q$) space. This can be thought of as a form of time-resolved correlation spectroscopy probing correlations between speckle patterns recorded at different time delays relative to the optical-excitation pulse. This correlation function has minimum value one for uncorrelated scattering patterns. A normalized quantity

$$\Delta C/\langle C(f,f)\rangle = \frac{\langle C(f,f) - C(t,f)\rangle}{\langle C(f,f)\rangle} \tag{2}$$

was also calculated, with $t$ denoting the transient speckle pattern collected at a specific time delay, $f$ denoting the final speckle pattern corresponding to $t = t_\infty$, and the angle brackets denoting an average across sample spots. This quantity corrects for the spot-to-spot variation in speckle contrast due to slight sample variations and x-ray beam quality differences (*Methods*).

In the following, we focus on the correlation between the transient intermediate structure at time $t$ and the final-state structure at $t = t_\infty$, as encoded in $C(t, f)$. This tracks the pathway of the non-equilibrium transient phase towards the final equilibrium state. Figure 2a shows the full two-time correlation function at two relative transient time delays averaged across sample locations, correlating each x-ray shot in the train with every other shot. The upper right quadrant correlates final shots with other final shots and is therefore significantly brighter, also showing that the final state has ceased to evolve by t=33 ms. The correlation of the transient shot with every other shot lies along the horizonal and vertical intercepts at $n = 0$ and grows in as time progresses, again exhibiting dynamics extending to ms timescales. Figure 2b quantitatively compares the time-dependent integrated intensity on the detector with the time-dependent correlation as computed from Eqns. 1, 2 in the ROI. As noted from a qualitative look (Fig. 1c), whereas the integrated intensity, a probe of the volume fraction of the new phase growing in, saturates on μs timescales, the correlation function grows in with a significantly more interesting temporal behavior, indicative of the more complex evolution of the dynamic heterogeneity associated with the new phase. The major feature is the long plateau that occurs between 500 ns and 100 μs, reflective of the lack of qualitative similarity between transient and final frames in Figure 1c. It is only after timescales of order 1 ms that the speckle patterns reach their final equilibrated state. This can also be seen by computing the fractional change in the two-time correlation function $\Delta C/\langle C(f,f)\rangle$ (Eqn. 2) (Fig. 2b). This follows a shape similar to that observed in the slow dynamics of heterogeneous soft-matter systems or supercooled liquids with a fast decay, long plateau, and then long-time fall-off.[25,26] The initial growth in $C(t,f)$ (equivalently, the initial decrease in $\Delta C/C$) corresponds to the emergence of the supercrystal satellite peak and its initial shift to its final position in $Q$ space (see Supplement for analysis of peak shape and location). Thus, although the dynamical response seems to be complete after a few μs as judged by standard crystallographic techniques, the correlation analysis reveals this to be incorrect.

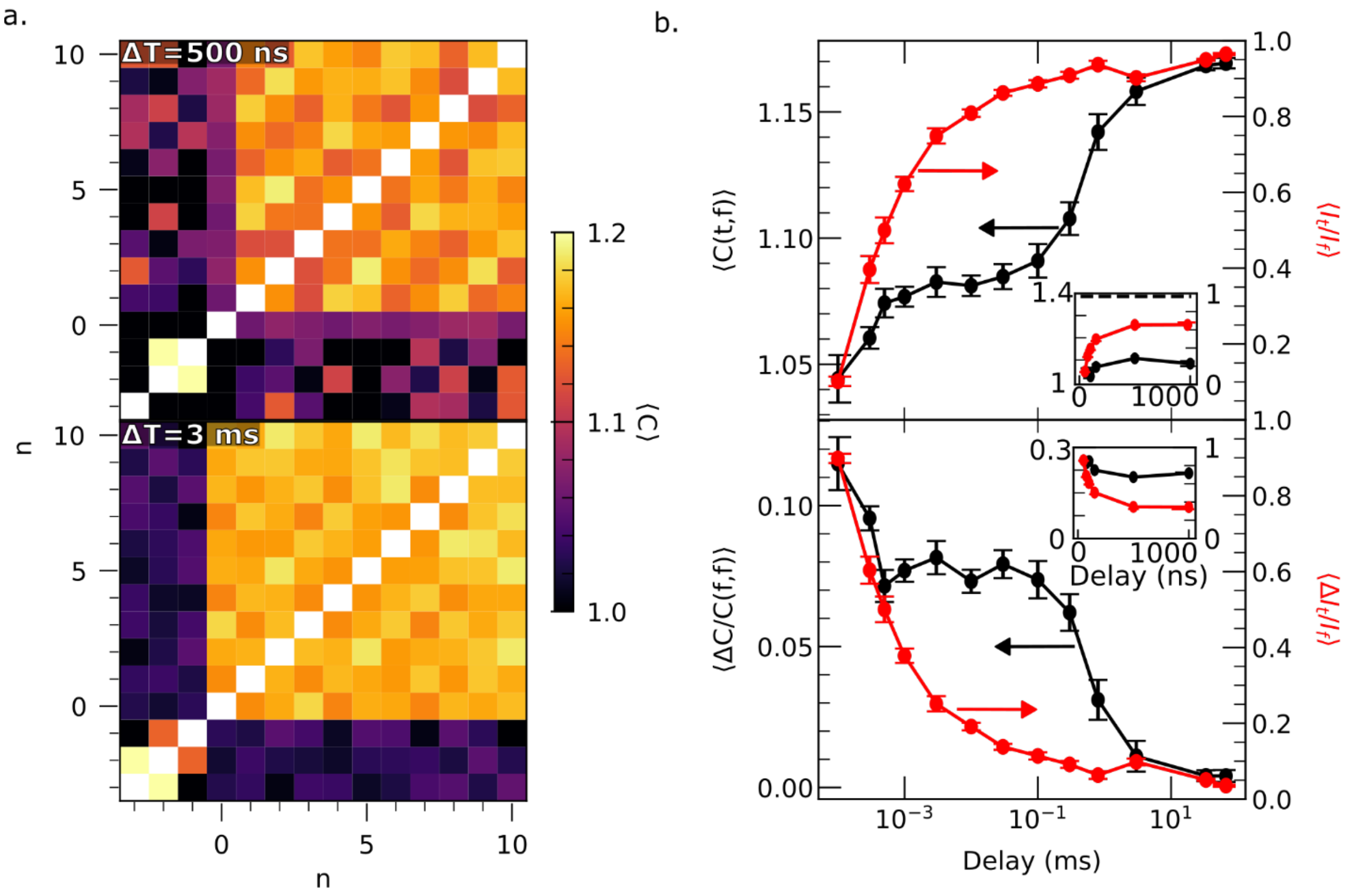

Fig. 2 a) Average two-time correlation plots at 500 ns (top) and 3 ms (bottom). The average was conducted across different sample locations pumped with the same pump-probe delay. b) Comparison of the normalized integrated intensity (red) to evolution of correlation function $\langle C(t,f)\rangle$ (black) (top). Comparison of the normalized differential intensity (red) to differential correlation normalized to the final correlation value (black) (bottom). Each time point was normalized to the average $C(f,f)$ at that time point. In both the top and bottom plots, the points at 33 ms and 66 ms were calculated by averaging the n = 1 and n = 2 rows of the two-time plots and are included to show the asymptotic value of $\langle C(t,f)\rangle$ and $\langle \Delta C/C(f,f)\rangle$. Insets show speckle correlation data at delays earlier than 1 μs.

To understand the microscopic processes responsible for the long-time non-equilibrium response, we first conducted a phase-field simulation of a PTO/STO superlattice as it transforms[27,28] (Fig. 3). Diffraction patterns and associated correlations functions using Eqn. 1 were computed for each simulated structure following reference [29] (*Methods*). First, the VSC phase nucleates at multiple locations and starts to expand, initiating the formation of the supercrystal phase with multiple grains. During this early time process, defects are generated at the boundary between the grains that do not anneal until much later in the simulation. Fig. 3a shows the real-space evolution from the simulation with additional movies in the Supplementary Information. The early-time process creates a local region of another intermediate structure with distinct periodicity from the original structure and the VSC phase as well as multiple sets of dislocation-like defects in the supercrystal region. This early process corresponds to an intensity increase of the supercrystal satellite peak that accounts for ~90% of its final equilibrium value but an increase in the correlation coefficient of the supercrystal satellite peak that accounts for only 50% of the final equilibrium value (Fig. 3b). The system then further undergoes a much longer process where the supercrystal defects gradually disappear. For the supercrystal satellite peak, ~10% of the final intensity and ~50% of the final correlation coefficient of the final value is gained within this second process. This gives rise to a two-component development of the correlation coefficient during the formation of the supercrystal (Fig. 3b), similar to that observed experimentally. As the process requires the transit and annihilation of defect pairs

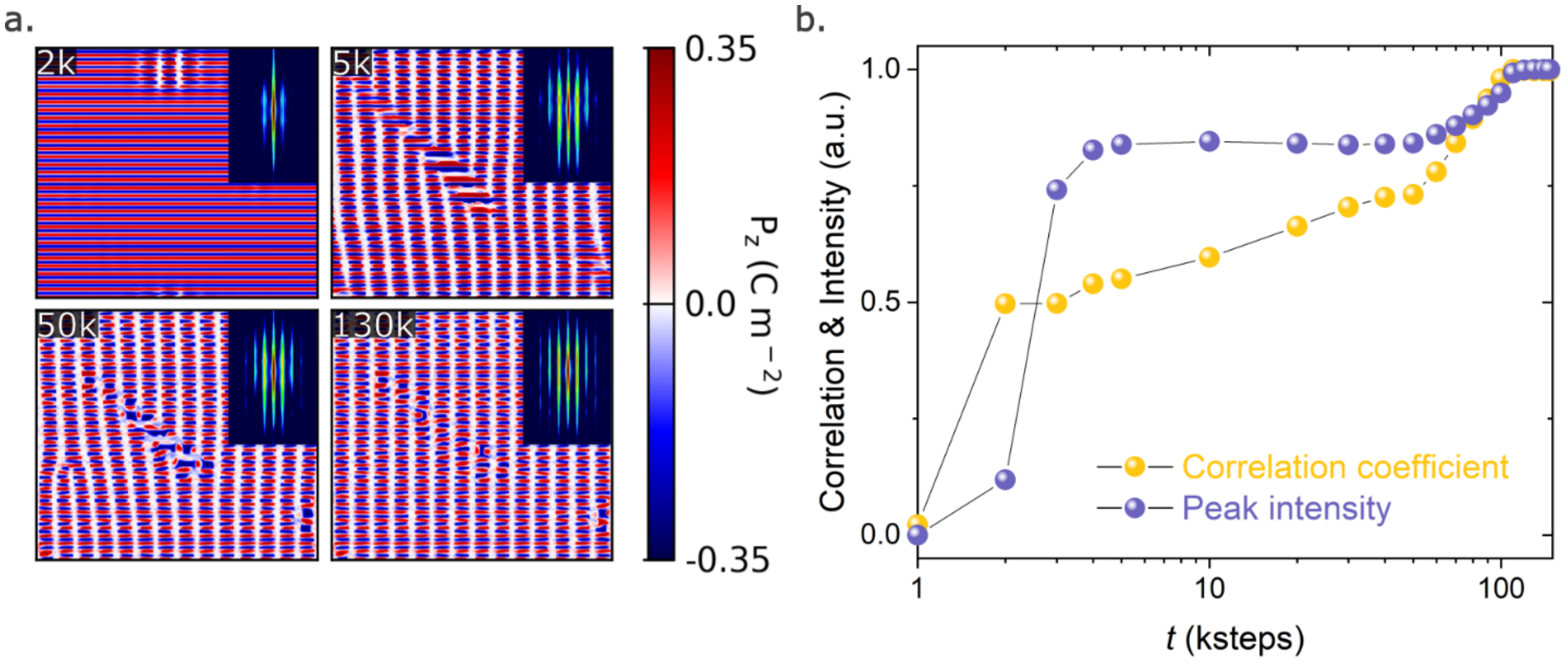

Fig. 3 a) Real-space phase-field simulation results of the heterogeneous evolution of the supercrystal growth as a function of time showing the z-component of the polarization vector in a 308 nm x 308 nm area (inset: corresponding computed diffraction pattern). b) Comparison of the first supercrystal satellite peak intensity (purple) to correlation function (yellow) as defined by equation 2 from data in a), over the entire first supercrystal satellite peak.

and grain boundaries, we expect the time duration to scale with the size of the simulation cell. Nevertheless, this shows the sensitivity of the correlation function to small-volume-fraction defects and domain boundaries as they annihilate and qualitatively captures the experimental results.

The phase-field simulations (Fig. 3) indicate that the different dynamics seen by the correlation function and the integrated intensity likely arise from the annihilation of defects at the boundaries of supercrystal-phase regions during the transformation. Qualitatively, such a sensitivity of the correlation function to such defect states and other types of heterogeneity can be understood from additional sensitivity in the correlation analysis to the diffuse scattering around the main Bragg peaks associated with nano/mesoscale ordering. An integration over an ROI in reciprocal space, as typically analyzed in a crystallographic experiment, integrates out the fine details of the diffuse scattering[30,31] which encode disorder and heterogeneity associated with the growth of the new phase, and thus provides information only on the volume fraction. In contrast, the two-time correlation function sensitively probes this, and is sensitive to the fine details in the scattering that reflect disorder, heterogeneity, and fluctuations within the growth of the new structure. In particular, as the domain walls evolve it is the high wave vector (off-Bragg-peak) components which change and these are picked up by the correlation approach. Thus, this approach enables one to obtain direct information about domain walls, the local-strain fields at the boundaries of the bubbles of new phase, and their dynamics.[17,32–35]

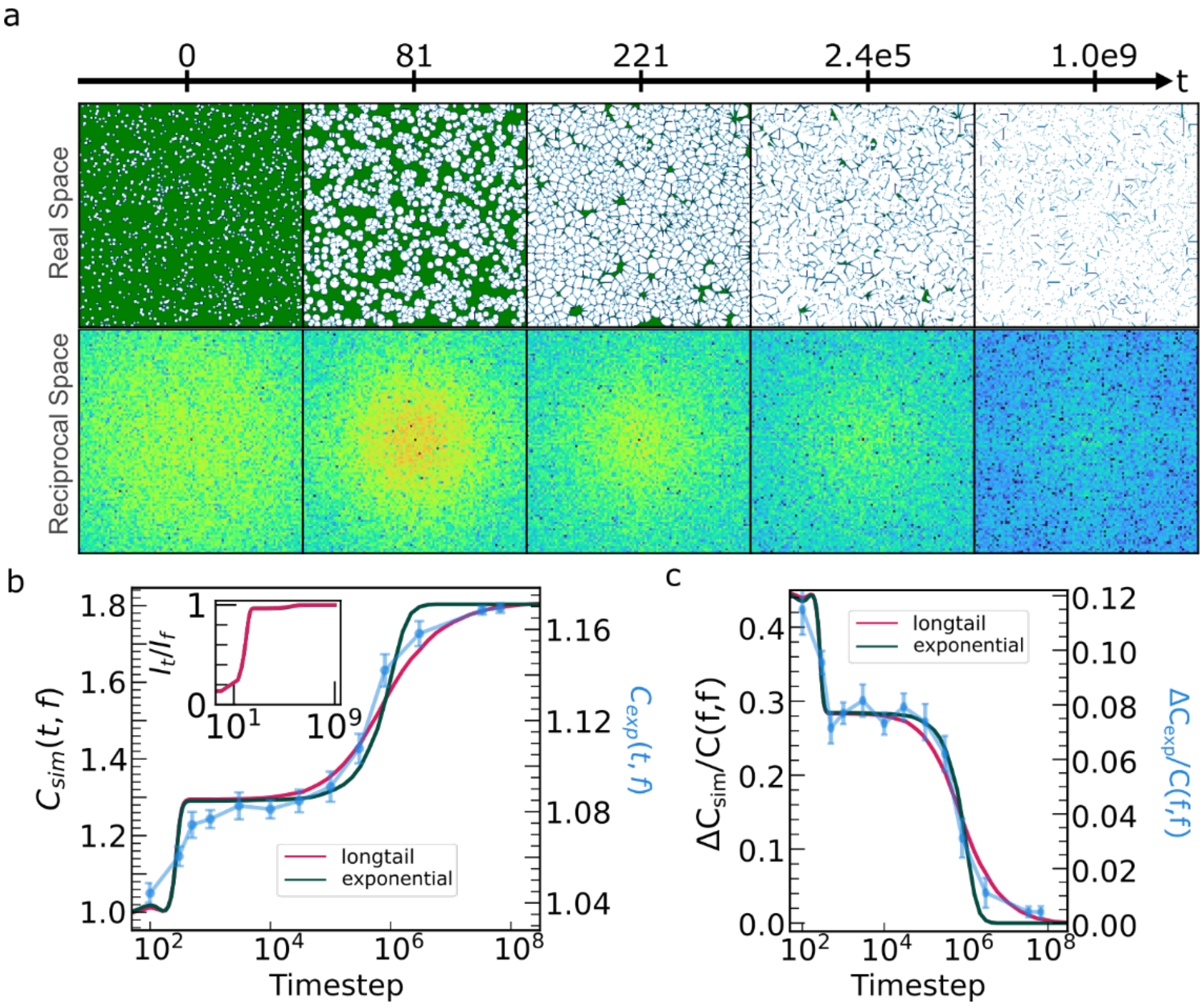

Fig. 4 a) (top row) Simulated nucleation and growth model snapshots showing isolated domains growing and impinging upon each other to form defect regions at the boundary between nuclei, followed by defect annihilation. (bottom row) The corresponding diffraction pattern for each image in the top row. b) Average over 250 simulations of the correlation function between the transient and final computed speckle pattern in an annular ROI. The generated defects are annihilated using two different waiting time distributions overlayed on the experimental raw correlation data. Inset shows the computed integrated intensity with negligible difference for the two waiting time distributions c) Overlayed average simulation and experimental normalized differential correlation data. Both b) and c) show the sensitivity in the long-time data to the waiting time distribution governing the annihilation of boundary defects.

To overcome the size limitation of the phase-field simulations, a series of phenomenological simulations were conducted, following approaches used in the modeling of spinodal decomposition[36]. A grid of points, representing small-area elements undergoing a phase transition was established with each point assigned either a value of 0 or 1 corresponding to the starting phase or the final supercrystal phase, respectively. A two-dimensional Fourier transform of this grid enables a simulation of a transmission mode XPCS measurement. The dynamics of nucleation and growth were simulated in this model by creating small regions of the final phase at random locations and allowing them to grow as disks until they impinge on neighboring regions of the final phase, creating many local boundary regions (*Methods*). Figure 4a shows the nucleation and growth process in one such simulation, together with the corresponding computed diffraction patterns. As expected, as the local regions expand the diffuse scattering qualitatively contracts inward, encoding information about the mesoscale structures and boundary regions.

Following an initial growth period, the boundary regions were allowed to reduce or annihilate at times sampled from a statistical waiting-time distribution[37] for the annihilation process, reflecting the stochastic jumps required to overcome an energy barrier associated with the merging/coalescence of domain walls. Similar models have been applied to domain-wall-based switching in ferroelectrics[38,39] and to anomalous transport in disordered materials.[40] In the simplest model, such distributions of waiting times arise from a heterogeneous distribution of energy barriers associated with these activated processes.[41] The waiting times assigned to each boundary was sampled from an exponential distribution decaying with timescale τ with $P(t;\tau) = \frac{1}{\tau}\exp\left(-\frac{t}{\tau}\right)$ and a 'longtail' distribution with $P(t;\tau,\alpha) = \frac{\alpha}{\tau}\frac{1}{\left(1+\frac{t}{\tau}\right)^{\alpha+1}}$ that decays at long times as $1/t^{\alpha}$. As a control, a uniform distribution was also considered and is shown in Supplemental Fig. S6. The exponential distribution is a prototypical memoryless distribution whereas the longtail distribution is associated with a scale-free fall-off and an infinite first moment (for $\alpha < 1$) which gives rise to long-time, non-equilibrium dynamics and effective sub-diffusive behavior of the domain walls.[40–42]

Three histograms for these assumed phenomenological waiting-time distributions and the computed correlation function and integrated intensity are shown in Fig. S6. A comparison and fit to experimental data is shown in Fig. 4b,c. Comparing the experimentally extracted correlation and normalized differential correlation curves shows the greatest similarity between the long-tailed distribution case and the experimental data (Fig. 4b,c). We obtain best qualitative agreement for $\alpha \approx 0.85, \tau = 1.5 \times 10^{5}$ in the case of the long-tailed distribution. This simulation, despite the simplicity of its assumptions, thus captures the central experimental observations, showing sensitivity to the dynamics of the domain-wall boundaries as they dynamically evolve, encoded in the time-dependent correlation function. Figure S6 shows fits for different values of $\alpha$, showing the sensitivity of the correlation function to power law tails and defining a new approach for extracting this parameter. The best fit value for $\alpha$ indeed gives rise to dynamics with infinite first moment, underlying the long-lived non-equilibrium response observed here. This sub-diffusive response may be understood in terms of the correlated interactions between different domains, as measured in prior slower timescale XPCS experiments probing jamming transitions.[43,44]

The extracted power-law-tail behavior of the waiting-time distribution for the domain-wall-annihilation process can in turn be related to an effective time-dependent growth of an average domain size, the central parameter within previously studied coarsening phenomena, phase-ordering kinetics, and other universal symmetry-breaking phenomena following an ultrafast quench.[4,17] We show in the Supplementary Information that in two dimensions, such a waiting-time distribution law gives rise to a domain-size growing algebraically in the scale-free manner $L(t) \sim t^{\alpha/2} = t^{0.43}$ in reasonable agreement with other theoretical work.[4,42,45] We note prior theoretical modeling of percolation dynamics in non-equilibrium coarsening models have indicated the presence of two time-constants; a fast and a slow one, as observed here in the correlation function.[46] We also observe that even simpler single nucleation models as described in the Supplementary Information can capture the observed delayed/two-step behavior of the correlation coefficient. These show enhanced sensitivity to the evolving heterogeneous strain field at the boundary of the domain, consistent with the above models.

In summary, this work defines new approaches for visualizing and understanding the ultrafast and heterogeneous non-equilibrium dynamics of mesoscale defects, and more broadly, the non-equilibrium pathways materials follow as they transform. Future work may enable combination of these approaches within device geometries under applied fields, with important technological applications. For example, ferroelectric device switching is mediated by nucleation and growth processes similar to those probed here. Tracking the pathways of such transitions via correlative scattering approaches would enable new types of calorimetry including sensitive probes of dissipation and entropy growth, which in turn define the ultimate speed limits[47] and energy costs for switching.[48] Expansion of x-ray free electron laser sources towards higher rep-rates should offer significant further improvements, potentially enabling direct real-space reconstruction of dynamically evolving defect pathways and their stochastic dynamics.

## Methods

*Experimental setup*

X-ray photon correlation spectroscopy experiments were conducted at the XCS hutch at the Linac Coherent Light Source (LCLS) at the SLAC National Accelerator Laboratory. The x-ray photon energy used for the diffraction experiments was 9.5 keV. X-ray pulses were fired at sample positions at a repetition rate of 30 Hz determined by a pulse-picker. Diffraction patterns were recorded using epix100 detectors located 8 meters away from the interaction point on the sample. Specifications can be found in reference [49]. The sample was pumped using a 50 fs 400 nm laser pulse produced by doubling an 800 nm laser pulse from a Ti:Sapphire amplifier.

*Materials Growth*

The $[(PbTiO_3)_n/(SrTiO_3)_n]_m$ ($n$- is the number of monolayers, $n$=16; m- is the period of the superlattice, $m$=8) superlattices were synthesized on single-crystalline $DyScO_3$ (110) substrates and via reflection high-energy electron diffraction (RHEED)-assisted pulsed-laser deposition (KrF laser). The $PbTiO_3$ and the top $SrTiO_3$ were grown at 610 °C in 100 mTorr oxygen pressure. For all materials, the laser fluence was 1.5 J/cm$^2$ with a repetition rate of 10 Hz. RHEED was used during the deposition to ensure the maintenance of a layer-by-layer growth mode for both the $PbTiO_3$ and $SrTiO_3$. The specular RHEED spot was used to monitor the RHEED oscillations. After deposition, the superlattices were annealed for 10 minutes in 50 Torr oxygen pressure to promote full oxidation and then cooled down to room temperature at that oxygen pressure.

*Averaging method for speckle analysis*

Each location on the probed sample produced a set of speckle patterns. One such set is shown in Fig. S1, with the speckle pattern taken as the transient image marked. The LCLS has a certain amount of shot-to-shot variation in the spatial location of the x-ray probe pulses, leading to slight differences in location of the observed satellite peaks between collected frames. To ensure a comparable ROI is chosen in correlation analysis, a fitting procedure was done to find the center of the final satellite peak. First, noise was reduced in the images by setting any pixel value less than 0.25 times the value measured when a single photon is detected to 0. The average final satellite peak was calculated at each sample position by averaging the final satellite images taken at the location each normalized to the incident intensity measurement taken for that image. A Lorentzian peak shape with a linear background (Eqn. S1) was fitted to the average vertical and horizontal profile to determine the center FWHM:

$$P(x; A, w\, x_0, m, b) = \frac{A * w}{(x - x_0)^2 + w^2} + m * x + b \quad \text{(S1)}$$

Once a center was determined, an ROI is chosen both in the transient image and the final images relative to the fitted center. The correlation function described in Eqn. 2 was then calculated over this ROI to produce the two-time plots shown in Fig 2a. The average $C(t, f)$ value was calculated by averaging the correlation value between the transient image and the final images at a single sample location, then averaging across locations with transient images with the same sample delay. The error bars were calculated as the standard error of the mean correlation value across locations. The same procedure was used to calculate the change in peak center and peak width, shown in Fig. S2a and S2b. The peak width and position were extracted from the fitted Lorentzian peaks.

*Multinuclear Phenomenological Modeling Procedure*

1000 nuclei were generated on a 500 x 500 cell grid. The center locations of the nuclei were randomly chosen so that no two nuclei overlapped centers. The nuclei were all generated before any evolution to approximate the experimental behavior of nucleation occurring effectively instantaneously upon laser excitation. Each nucleus was assigned a unique number in the range [0,1000). For each cell, the ratio

$$r = (d_2 - d_1)/(d_2 + d_1) \quad \text{(S2)}$$

was calculated, where $d_1$ is the distance between the cell and the closest nucleus, and $d_2$ is the distance between the cell and the 2nd closest nucleus center. Each cell (grid point) was assigned a unique index based on the 2 nearest nuclei for the purpose of assigning border regions to pairs of adjacent nuclei. The equation for generating the unique index is based on the Cantor pairing function $\pi(k_1, k_2)$.

$$m(n_1, n_2) = \pi(|n_1 - n_2|, n_1 + n_2) = ((n_1 + n_2) + |n_1 - n_2|) * ((n_1 + n_2) + |n_1 - n_2| + 1)/2 + (n_1 + n_2) \quad \text{(S3)}$$

where $n_1$ and $n_2$ are the indices of the individual nuclei and $m$ is the index assigned to the cell. To determine border regions, cells with r < 0.05 were permanently cell to 0. Initial nuclei were generated by creating regions between 1 and 3 cells in radius around the randomly generated centers. For 1000 timesteps, a number of randomly chosen nuclei determined by sampling a Poisson distribution with mean of 63.2 at each timestep were allowed to grow 1 pixel in radius. The speckle pattern of the domain structure during this growth phase was calculated using the Fast Fourier Transform (FFT) at 51 evenly spaced timesteps between 1 and 1001. The border regions were allowed to reduce or annihilate for $10^9$ timesteps. Each border region with a unique index m was assigned a waiting time sampled from a uniform distribution, an exponential distribution (Eqn. 3a), or a long tail distribution (Eqn. 3b). When the timestep of the simulation reached the waiting time of any given border region, that region was annealed by instantaneously changing to a region with effective r =0.01. This reduced some border regions' thickness and eliminated others. The initial border regions and the final border regions are shown in Fig. 4a and 4b. The border defect regions are colored in Fig. 4a and 4b according to the label $m$ given to them according to Eqn. S3. The speckle pattern during the defect reduction phase was calculated at 51 logarithmically spaced timesteps. All correlation functions were calculated for annular ROIs of inner radius 2 and outer radius 10. An example two-time of this system is shown in Fig. S5. The correlation plot shown in Fig. 3 is the 102nd row (equivalently 102nd column).

*Fitting multinuclear simulation results*

To find appropriate values of the waiting time distribution parameters (e.g., $\tau$ for the exponential distribution and $(\alpha, \tau)$ for the long tail distribution), a fitting procedure was performed. Each timestep in the simulation was taken to be 1 ns. For the exponential and long tail distribution, 10 simulations using the same procedure as above were performed except that the growth steps were not performed, and the growth was assumed to be completely done by 3 µs. The resulting $\Delta C/C(f,f)$ curves were averaged. An example set of C and intensity curves for different values of $\alpha$ for $\tau = 1.5 * 10^5$ are shown in Fig. S6. The $\Delta C/C(f,f)$ curve was normalized to decay from 1 to 0 and compared to the experimental data that was normalized such at that $\Delta C/C(f,f)$ at 3 µs was 1. The residual between these two curves calculated at the experimental delay points was minimized using a non-linear minimizer until the values reported in the main text were found.

*Phase-Field Simulation Method*

The simulation system of the $PbTiO_3/SrTiO_3$ superlattice contains 2 $PbTiO_3$ layers and 2 $SrTiO_3$ layers, where each layer has a thickness of 4.8nm, equivalent to 12 unit cell lengths of the perovskite ($PbTiO_3$ or $SrTiO_3$) crystal lattice. The in-plane size of the system is chosen as $308\ \mathrm{nm} \times 308\ \mathrm{nm}$. The method and parameters for computing the scattering intensity of the structural domains are the same as performed by Yang *et al*.[29] An animation showing a 3D depiction of the $P_z$ as a function of timestep is shown in Fig. S7.

Two simulations were performed with slightly different values of epitaxial strain. With a strain of 0.2%, an intermediate phase appears during the phase transition, but alternate pathways appear with different values of strain (Fig. S8). Despite the difference in pathway, a two-step process nonetheless is apparent, indicating that the correlation function is still a sensitive measure of the defect formation and annihilation process which occurs generally in this formation process.

*Mononuclear Diffraction Simulation*

A 3$^{rd}$ set of simulations further demonstrated the sensitivity of the correlation function to the dynamics and structure of a domain wall during a phase transition. In this model, shown in Fig. S9, we assume a 2D periodic lattice of atoms and nucleate a new phase at the center of the grid, defined by a structure with slightly smaller lattice parameter. We imagine this new phase grows in by expanding radially with time until the old phase is subsumed by the new one. In the picture shown on the left of Fig. S9a, we depict a snapshot of this transformation for the case of an atomically sharp domain wall separating the new phase from the old one. In contrast, the right figure of S5a shows the corresponding snapshot for a diffuse domain wall where a gaussian strain gradient separates the new phase from the old one. With these two charge density snapshots, one may then directly compute the corresponding diffraction patterns at each step in the simulation by Fourier-transforming the charge density with resulting snapshots shown in Fig. S9b. Finally, from these diffraction patterns we may extract the time/radial dependence of the intensity and correlation function. For a sharp domain wall, the intensity and the correlation function track each other approximately whereas for for the more diffuse case we find that the correlation function evolves significantly more slowly than the integrated intensity, showing as observed experimentally, that the correlation function more sensitively probes the structure of the domain wall.

**Acknowledgements.** This work was primarily supported by U.S. Department of Energy, Office of Science, Office of Basic Energy Sciences, under Award Number DE-SC-0012375. The single-shot XPCS experiment, development of the data analysis, and forward modeling was partially supported by the U.S. Department of Energy, Office of Science, Office of Basic Energy Sciences, Materials Science and Engineering Division. Use of the Linac Coherent Light Source (LCLS), SLAC National Accelerator Laboratory, is supported by the U.S. Department of Energy, Office of Science, Office of Basic Energy Sciences under Contract No. DE-AC02-76SF00515. The phase-field effort is primarily supported by the US Department of Energy, Office of Science, Basic Energy Sciences, under Award Number DE-SC0020145 as part of the Computational Materials Sciences Program. J.J.T acknowledges the U.S. Department of Energy, Office of Science, Basic Energy Sciences, Materials Sciences and Engineering Division under Contract No. DE-AC02-76SF00515 under the NEMM program MSMAG. P.M. and L.W.M. acknowledge additional support from the Army Research Office under the ETHOS MURI via cooperative agreement W911NF-21-2-0162 for the production of the oxide superlattices.

**Author contributions.** Experiments were carried out by A.M., V.A.S., H. Wang, Q.L.N., S.S., M.C., Y.S., J.T., J.W.F., H.W., D.Z., Y.C., and A.M.L. S.D., P.M., and E.D. synthesized the samples with support from L.W.M. T.Y., M.Z., and L.-Q. Chen led the phase-field modeling efforts. Data analysis was carried out by A.M., H.Z., and Y.C. with support from A.M.L., D.Z., and V.G. A.M. carried out the phenomenological simulations shown in Fig. 4 with support from A.M.L., Y.C., and V.G. All authors contributed to the writing of the manuscript.

**Competing interests.** The authors declare no competing interests.

**Supplementary**

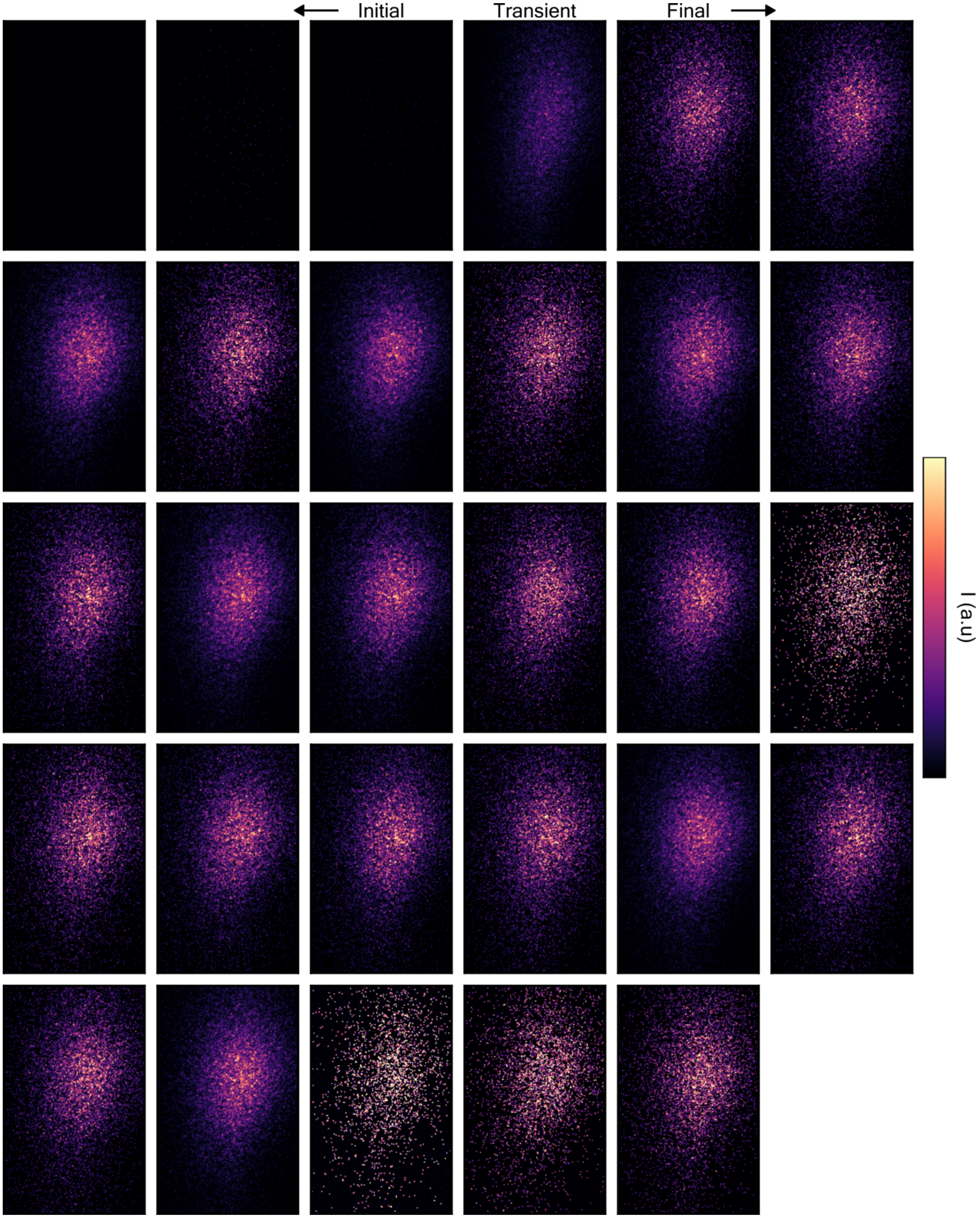

Fig. S1. A complete set of supercrystal satellite speckle patterns collected at a single location of the sample with a pump-probe delay of 1 μs. These diffraction peaks were normalized to i0 using a diode monitor. The three left-most images of the top row were collected before optically pumping the PTO/STO superlattice. The image labeled "Transient" is the supercrystal satellite peak at 1 μs after the optical pump. The subsequent images were collected with a repetition rate of 30 Hz after the optical pump. The image to the right of the transient therefore is the speckle pattern at this location at 33 ms after the optical pump, and the right-most image of the second row is the speckle pattern at 66 ms.

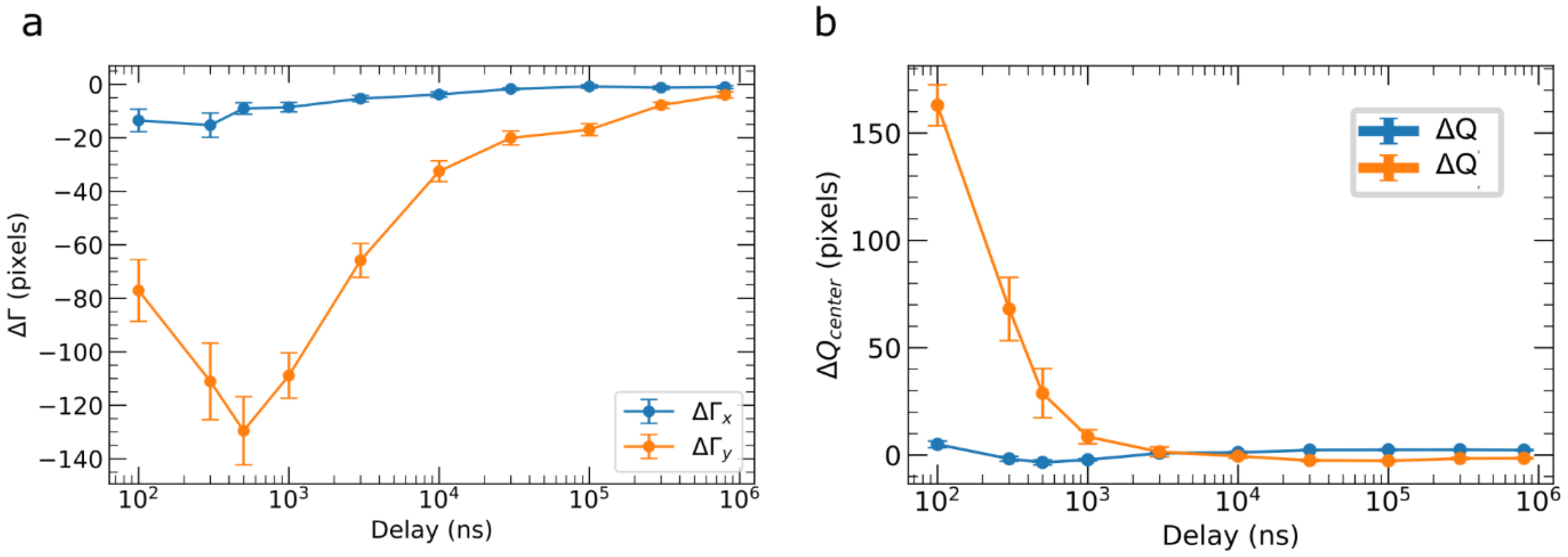

Fig. S2 a) The average change in supercrystal satellite peak width in both detector directions. The width along x (y) is the width of the peak fitted to a 1D Lorentzian after integration along the y (x) direction. The width in the x direction is the width in 2θ. The width in the y direction is the width in χ. b) The average change in peak center location in both detector directions. The characteristics of the peak evolve much more quickly than the speckle correlation function, indicating that mesoscale dynamics of the sample resulting in speckle evolution occurs on different timescales.

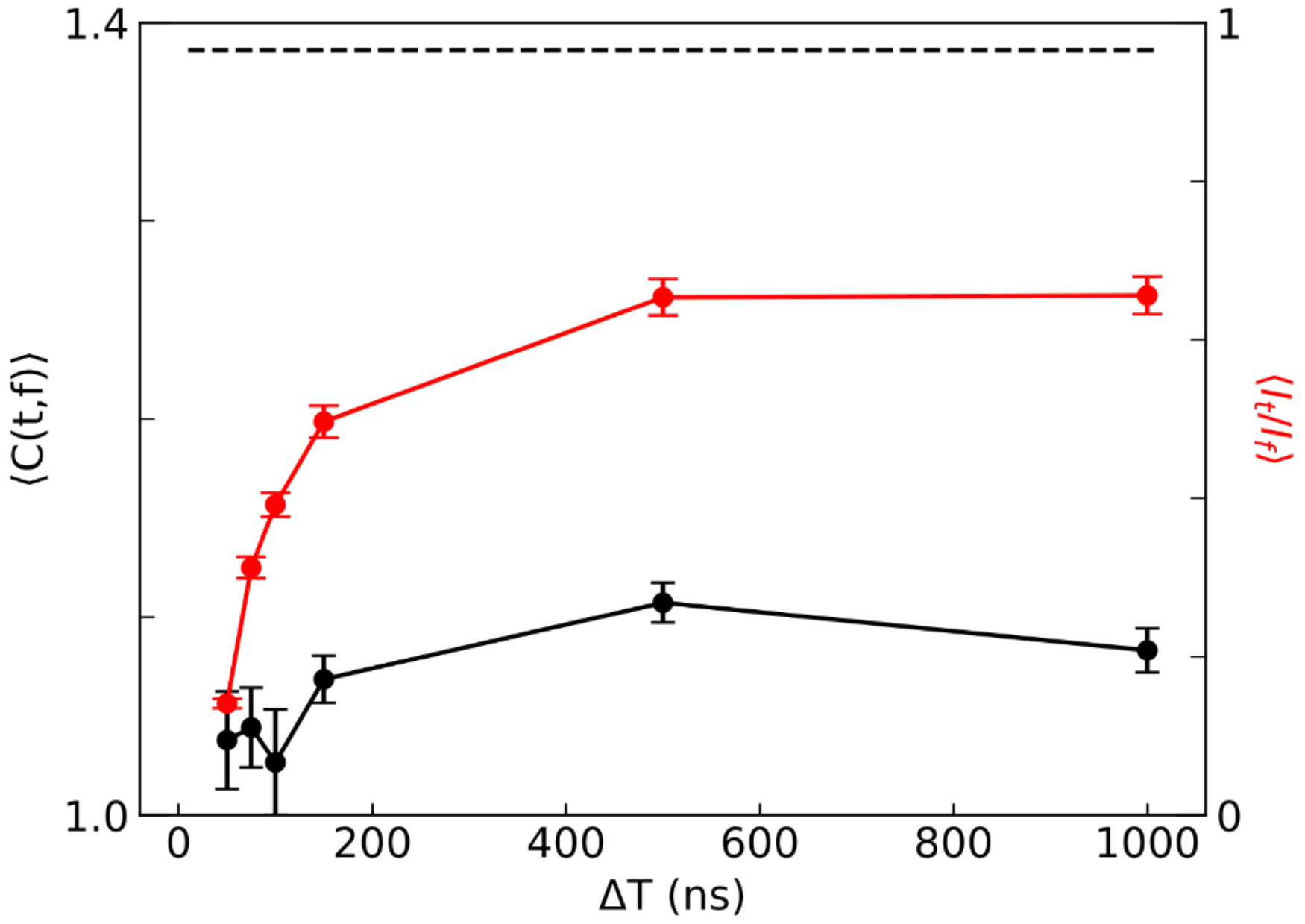

Fig. S3. The average correlation function calculated using Eqn. 1 from the main text at early times up to 1 μs. This dataset was collected with slightly different beam coherence and other experimental conditions. The difference in experimental parameters resulted in a higher final value for C (indicated by the dashed line), and slightly different intensity growth characteristics. The value of C up to 1 μs has not approached the final value of C and therefore speckle patterns at the longer delay values seen in the main text were collected.

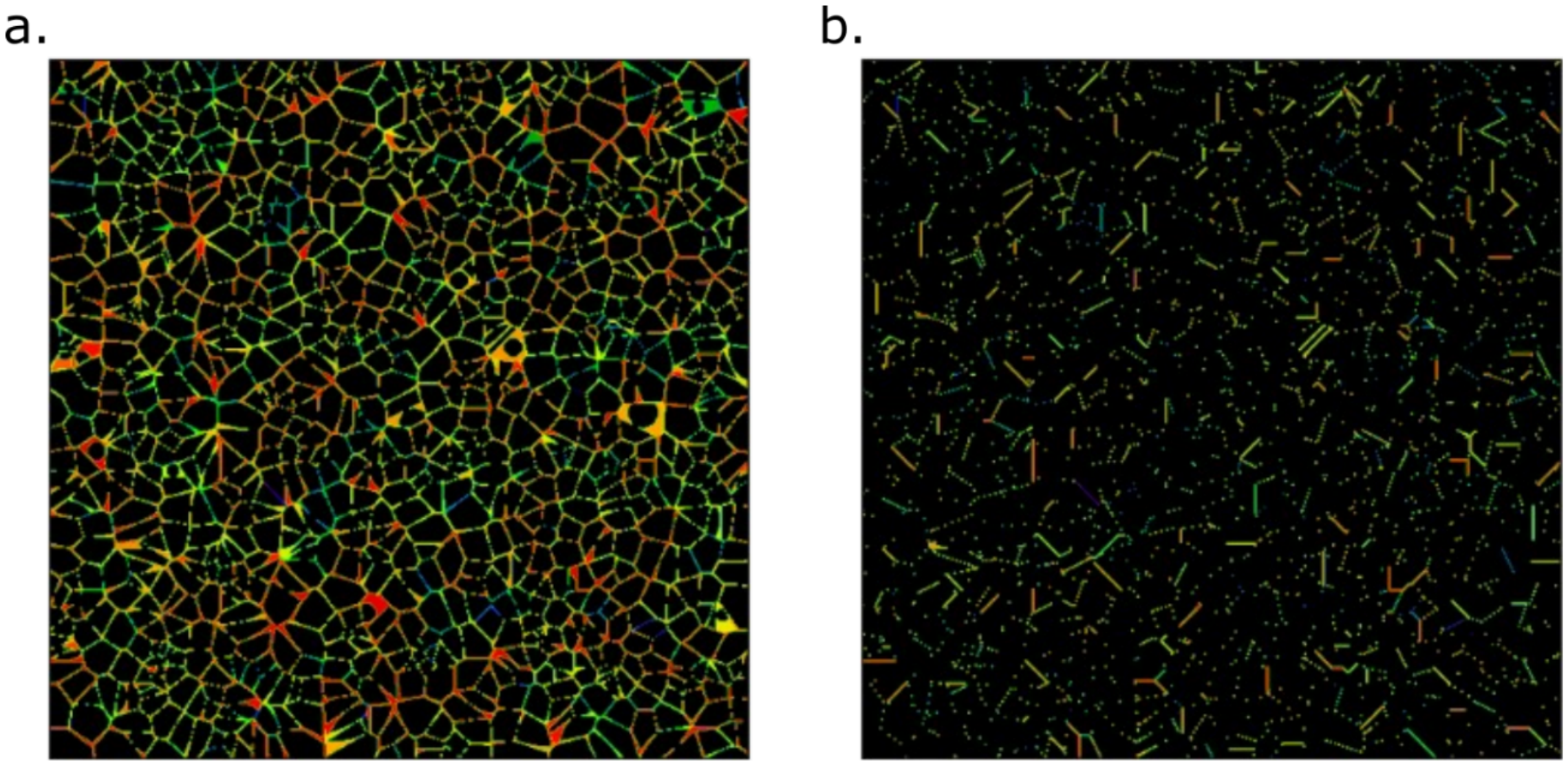

Fig. S4. The precalculated defect regions for the multinuclear simulation. Each border region is assigned a unique number based on the pair of nuclei that originally resulted in its formation. a) The defect regions with numerical labels after the growth phase with initial r=0.05. b) The defect regions with numerical labels after the anneal phase with final r=0.01. The anneal occurred according to different waiting time distributions.

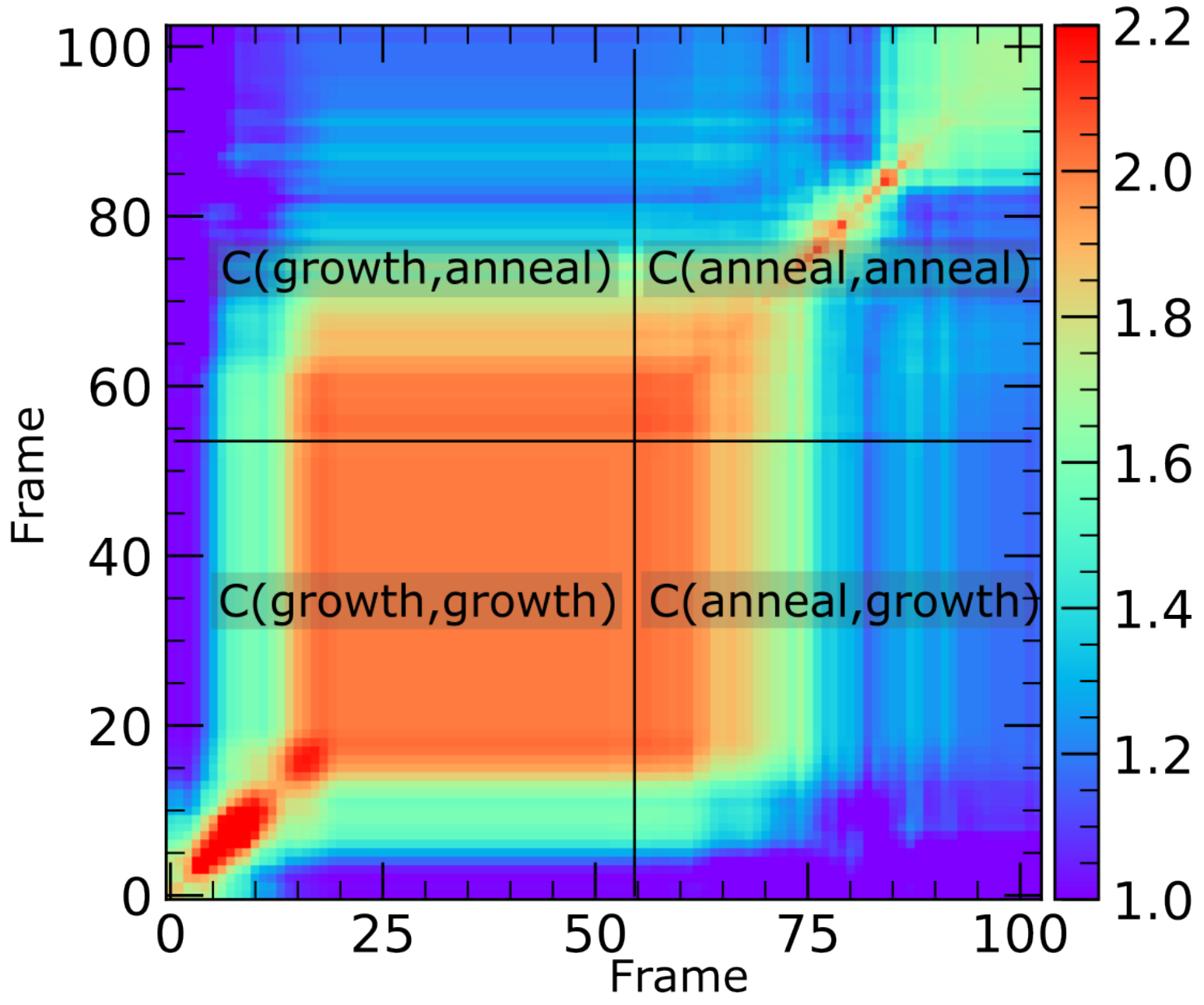

Fig. S5. A two-time function calculated for an annular region in the speckle pattern calculated in the simulation model presented in Fig 4 in the main text. A single simulation was run with a long-tailed distribution function with α=0.75 and τ=1.5E5. Roughly 50 speckle patterns were calculated during the growth phase (when nuclei were formed and grown) and the anneal phase (when defects were reduced or annihilated) each. The bottom left labeled quadrant represents the correlation between the speckle patterns calculated during the growth phase with other speckle patterns calculated during the growth phase. The top right quadrant represents the correlation between the speckle patterns calculated during the anneal phase with other speckle patterns calculated during the anneal phase. The bottom right and top left quadrants represent the correlation between the frames in the growth phase and the frames in the anneal phase. The C curves in the main text are extracted from the 100th row and averaged over 250 instances of the simulation.

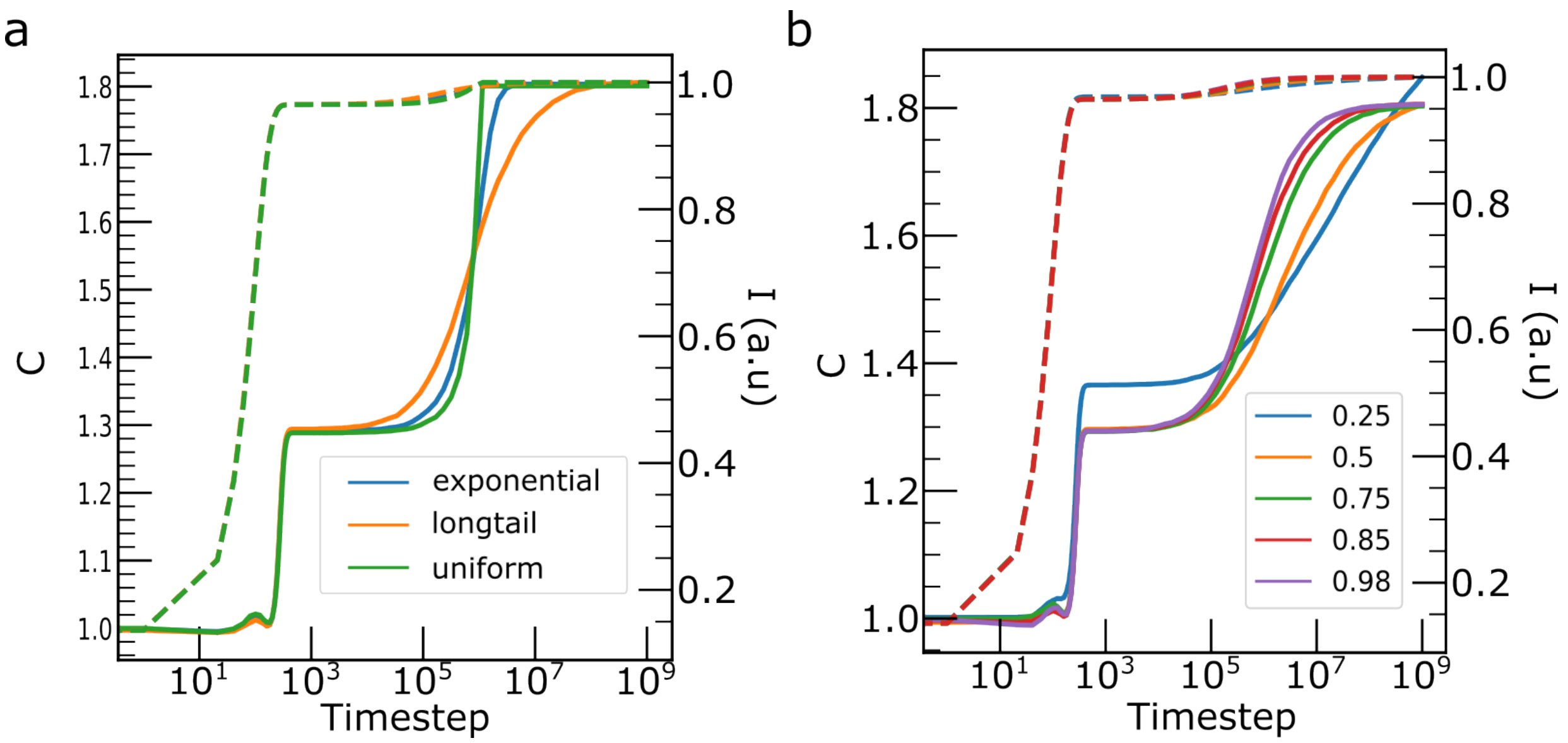

Fig. S6. a) Comparing C curves from simulations using three different waiting time distributions showing the control of qualitative shape b)A comparison of the effect of using different values of $\alpha$ in the long tail defect waiting time distribution on C (solid lines) and intensity (dashed lines). The rise of the C as a function of timestep is drastically changed as $\alpha$ is varied between 0 and 1. Assigning waiting times according to the long tail distribution is therefore a powerful method of fitting the phenomenological model in this paper to the data collected at the LCLS.

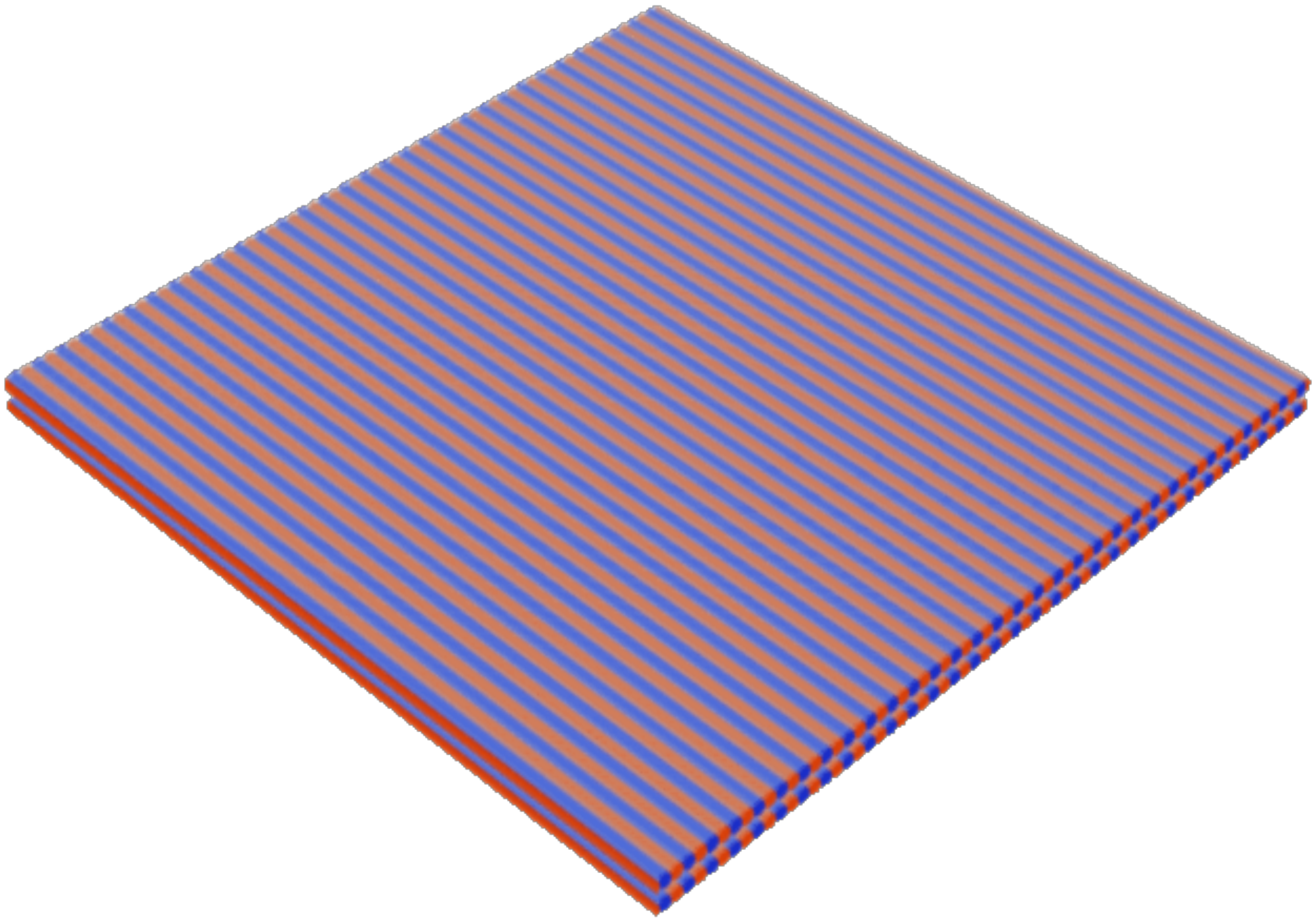

Fig. S7. An animation of the evolution of the z component of polarization (parallel to the pseudo-cubic c axis) over the course of the phase field simulation presented in Fig. 3 of the main text.

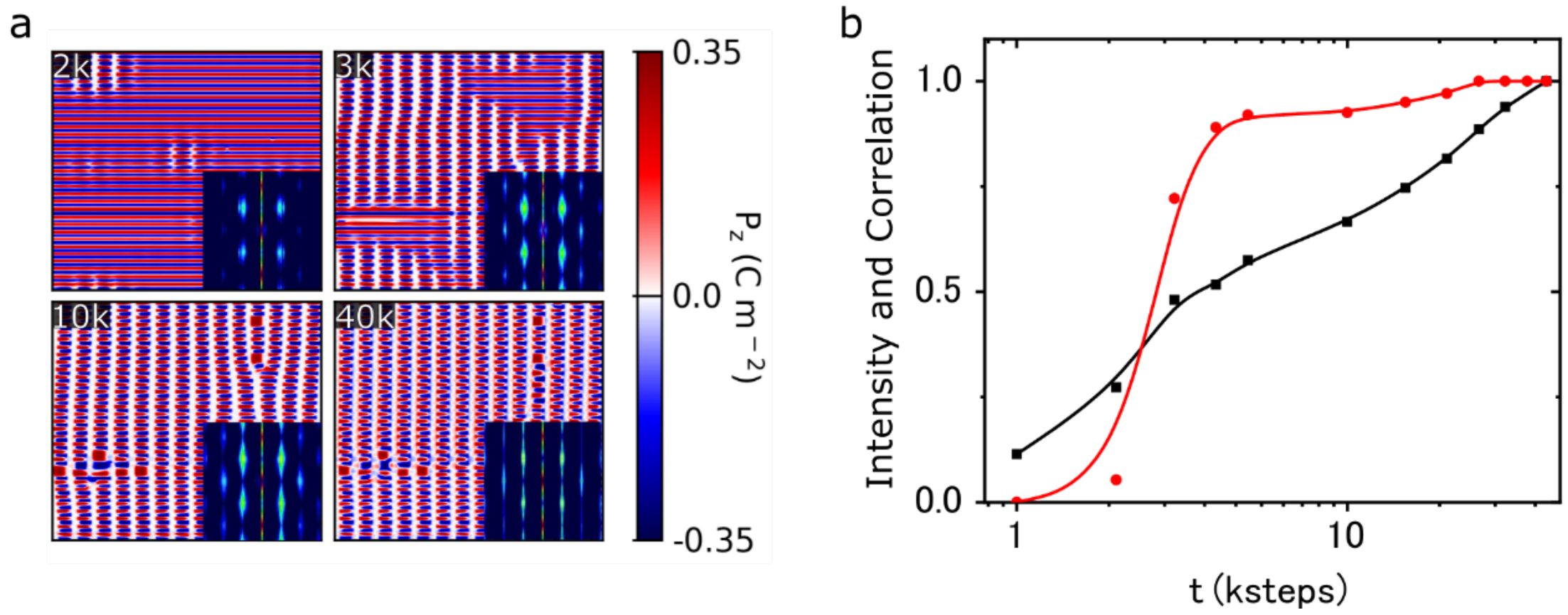

Fig. S8. a) Real-space phase-field modeling of the heterogeneous evolution of the supercrystal growth as a function of timestep (inset: corresponding computed diffraction pattern). b) Comparison of integrated intensity (red) to correlation function (black) as defined by equation 2 from data in a) . An epitaxial strain of 0.0% is taken.

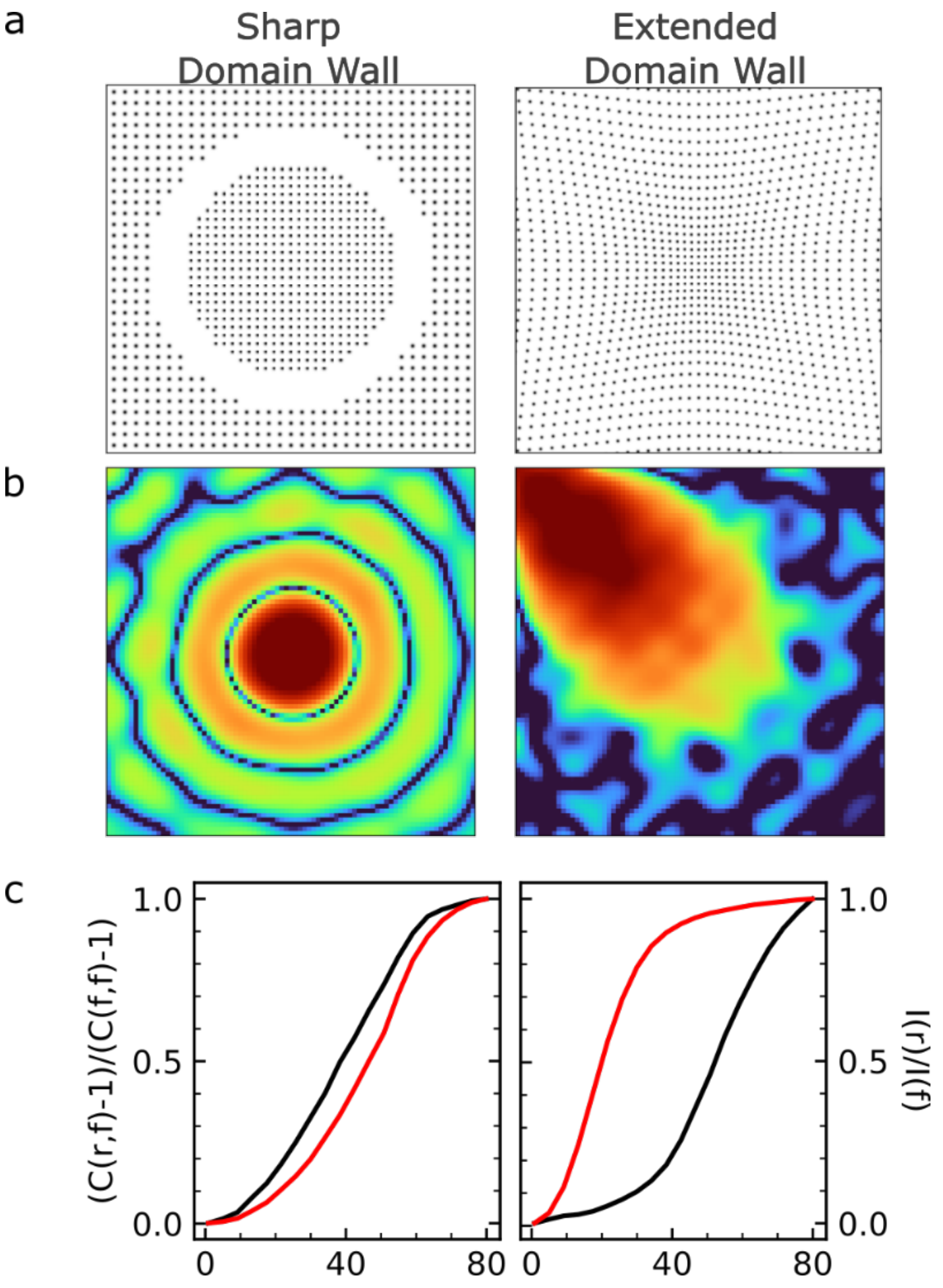

Fig. S9. a) Charge density for the case of a very sharp domain wall between the new phase and the old phase (left) and an extended domain wall (right). b) The diffraction pattern around the position of the peak associated with the new phase calculated by performing an FFT in the case of the sharp domain wall (left) and extended domain wall (right). c) Normalized correlation function for the sharp domain wall diffraction pattern ROI (left) and the extended diffraction pattern ROI (right). The delay between the correlation and the intensity curves in this case is changed dramatically by the domain wall types, therefore quantifying the sensitivity of the diffuse scattering to the domain wall.

## Extraction of domain size scaling law from waiting time distribution

We consider the 1D and 2D cases separately. In 1D, consider a a chain of N domains of unit size each initially separated by a domain wall.There are N-1 walls. We consider for simplicity the case N=8 initially as shown in Fig. S10 and then generalize to arbitrary N. At time zero the average domain size <R(0)>=8/8=1. After one annihilation event (t=1 in the figure), a wall is removed at random say at the 3th position. There are now 7 domains with <R(1)>=(1+1+2+1+1+1+1)/7=8/7. At time step 2 another wall is removed at say position 5 such that <R(2)>=(1+1+2+2+1+1)/6=8/6. This sequence continues until at the final time step, <R(7)>=8. It is clear that the average domain size progression does not depend on the order in which the walls are removed.

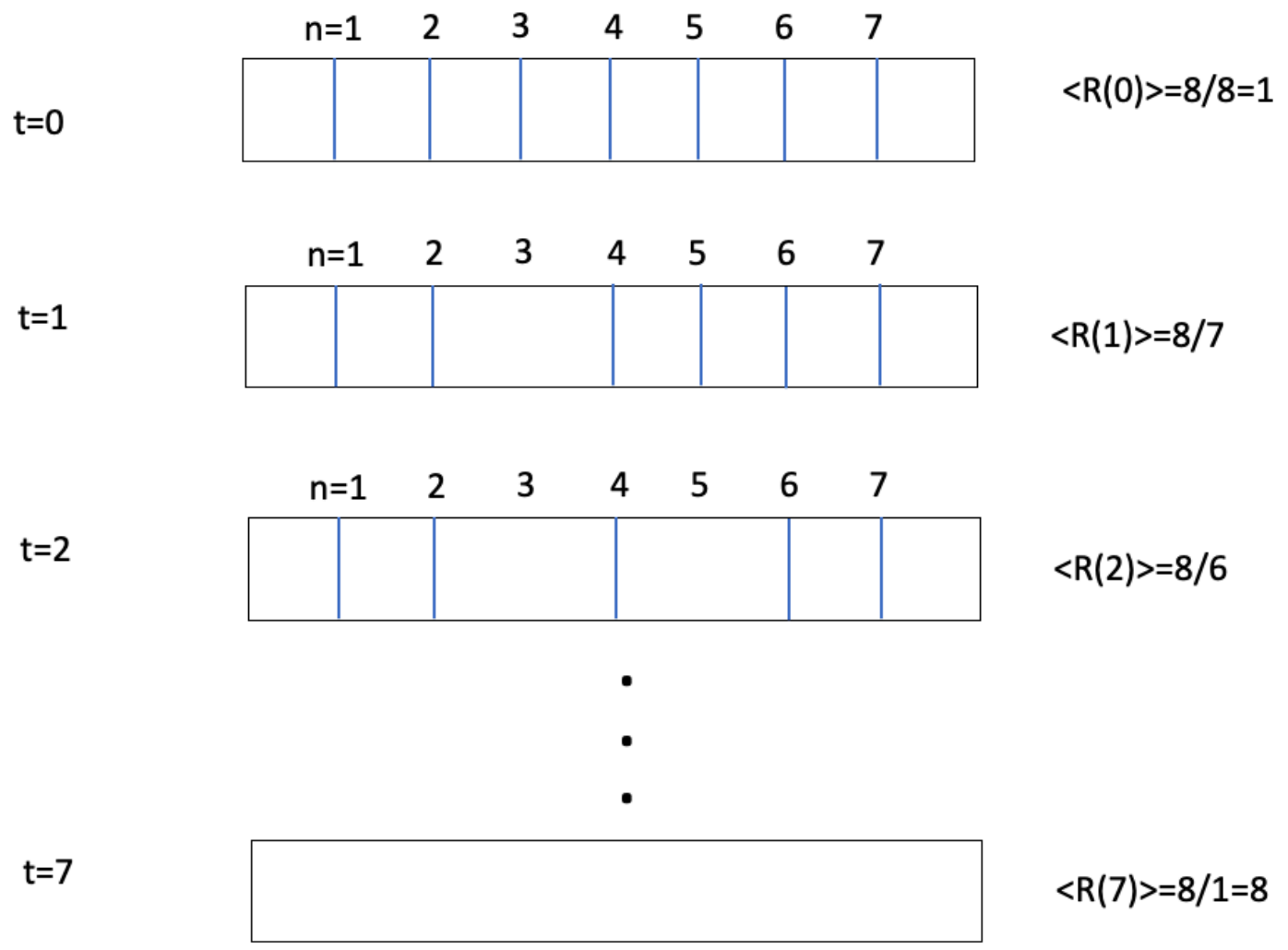

Fig. S10. 1D model for calculation of time-dependent domain size.

For N domains and N-1 walls at discrete time step t, $<R(t)>=N/(N-t)$. If instead the walls are removed following a waiting time probability distribution P(t) then the number of annihilated walls at time t is $(N-1)\int_0^t P(t')dt'$ and thus

$$.< R(t) >= \frac{N}{N-(N-1)\int_0^t P(t')dt'}$$

This equals 1 at t=0 and approaches N as $t \to \infty$ as expected. For $P(t)=\frac{\alpha}{\tau}\frac{1}{\left(1+\frac{t}{\tau}\right)^{\alpha+1}}$ as found in the main text, we then find

$$< R(t) >= \frac{N}{1+\frac{N-1}{\left(1+\frac{t}{\tau}\right)^{\alpha}}}$$

In the large N limit, $<R(t)>=\left(1+\frac{t}{\tau}\right)^{\alpha}$ and then in the large t limit, $< R(t) \sim t^{\alpha}$.

In 2D the same argument applies except instead to the average domain area, as shown in Figure S11.

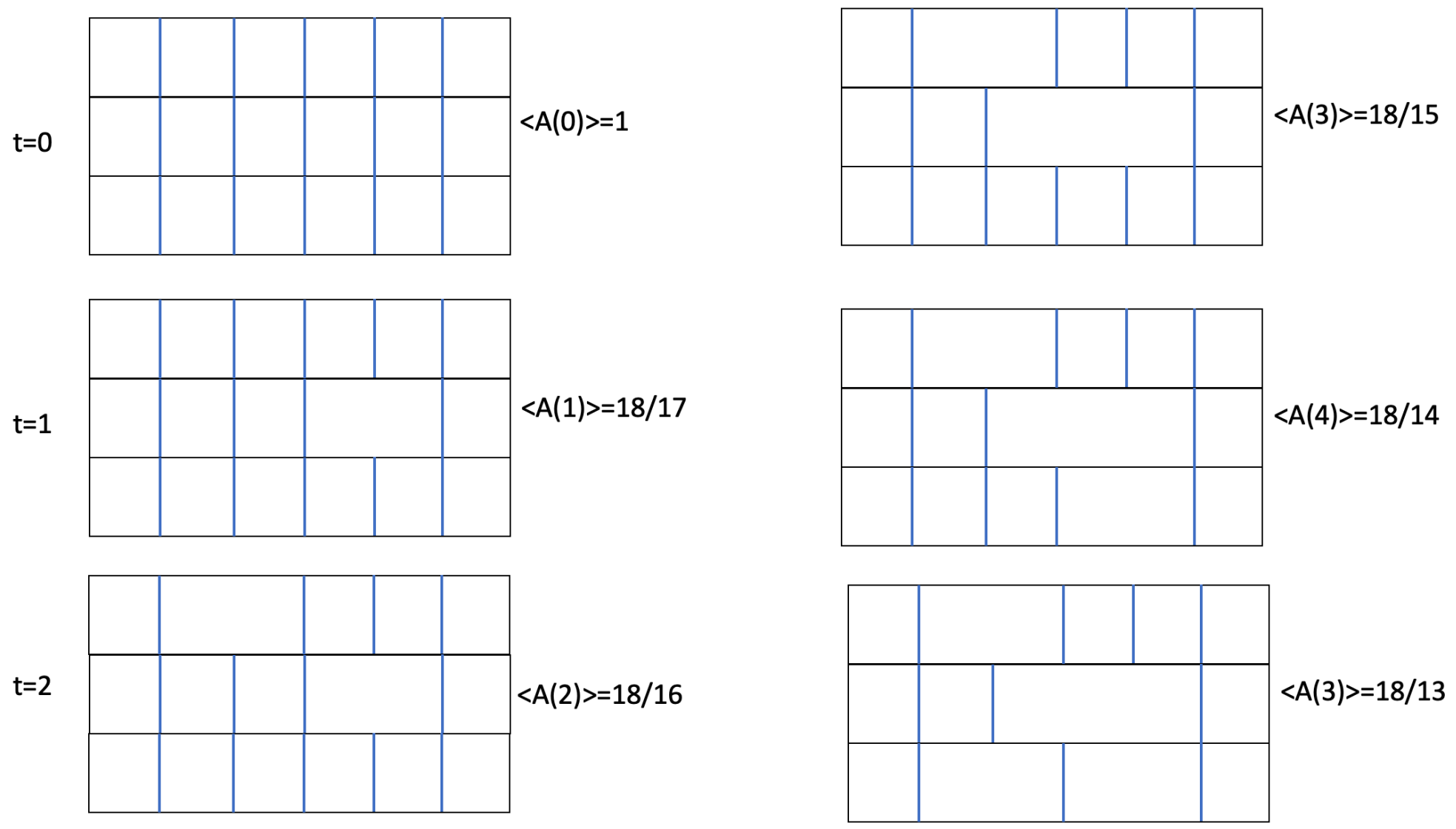

Fig. S11. 2D model for calculation of time-dependent domain size.

Thus $< A(t) > \sim t^{\alpha}$ and in 2D, $< R(t) >= t^{\alpha/2}$ as stated in the main text.